# Nightfall: Can Kalgash Exist?

Smaran Deshmukh[1] & Jayant Murthy[2]

## Abstract

We investigate the imaginary world of Kalgash, a planetary system based on the novel 'Nightfall' (Asimov & Silverberg, 1991). The system consists of a planet, a moon and an astonishing six suns. The six stars cause the wider universe to be invisible to the inhabitants of the planet. The author explores the consequences of an eclipse and the resulting darkness which the Kalgash people experience for the first time. Our task is to verify if this system is feasible, from the duration of the eclipse, the 'invisibility' of the universe to the complex orbital dynamics.

## 1. Introduction

Nightfall was first written as a short story by Isaac Asimov in 1941 (Asimov, 1941) and was then expanded into a full-length novel by Asimov & Silverberg in 1990 (Asimov & Silverberg, 1991). The story is set in a world with 6 suns which conspire such that at least one is in the sky at all times ensuring that the planet and its inhabitants never see darkness. Due to a chance combination of events, different groups of scientists come to the conclusion that:

1. There are cycles with a regular period of 2049 years in the planetary civilization such that the civilization at the time is destroyed by fire.
2. These cycles correspond to eclipses by a previously unknown moon which results in a period of complete darkness lasting for several hours. Because the inhabitants of the planet had never experienced darkness in recorded history, this drove them to madness and destroyed their civilization.

The short story focuses on the events leading up to the eclipse while the full length novel continues the story to the rebuilding of civilization by a religious society.

Asimov was one of the most influential writers of the Golden Age of Science Fiction and Nightfall, itself, is often ranked as the best short story of the genre (Sagan, 1992). Nightfall is an example of what was known as "hard science fiction" which aimed to have a realistic scientific base. However, Asimov, himself, was not very concerned with scientific detail as long as the events were plausible and Nightfall was written with the intent (after prodding by the legendary publisher John Campbell) to explore the consequences of the Ralph Waldo Emerson quotation "*If the stars should appear one night in a thousand years, how would men believe and adore, and preserve for many generations the remembrance of the city of God!*". Asimov and Campbell thought that the sight of the stars would rather drive men mad.

As a result, Asimov imagined a system which met the requirements of plausibility but certainly did not actually work out the mathematical details to see if the system was dynamically stable or if the details of the orbits would work out to give the necessary period of darkness. When Asimov conceived of Nightfall, no systems outside the Solar System were known; Kepler is finding more and more systems, many much different from our own including many with multiple suns (Orosz, et al., 2012). We were inspired by this to work out the details of the stellar system in Nightfall. We do this as a tribute to Asimov whose work inspired so many people (including one of the present authors) to learn more about the world around us.

[1]smarandeshmukh@gmail.com
[2]Indian Institute of Astrophysics, Bangalore 506034, India (murthy@iiap.res.in)

## 2. SUMMARY OF THE STORY

The short story and the later novel are both set in the same world albeit with minor differences in names and characters. However, the book contains much more detail and we have used names and constraints from the book for the sake of consistency. The story is set on the planet Kalgash which orbits the Sun-like star Onos at a distance of about 1.2 AU (1.2 times the Earth-Sun distance). There are five other stars (Dovim, Tano, Sitha, Trey, Patru) in this system and, as a result, Kalgash has the normal day/night cycle with respect to Onos but with one or more of the other stars always in the sky. The sky is bright enough that no other stars are seen and the inhabitants have no knowledge of the wider Universe. Presumably the motions in this system are complicated enough that the Universal Law of Gravitation was only discovered a few years before the events in the story. The entire extent of the system is 110 light minutes (13.2 AU) and, as far as the inhabitants of the planet were concerned, included all of Creation.

The story (in the novel) begins with an archaeological dig in which the scientists discover that there have been many previous civilizations which regularly destroyed themselves by fire, in accordance with the religious texts of a major cult. At about the same time, a group of astronomers found that the observed motions of the system did not match the predictions of the Law of Gravitation and postulated an additional invisible moon of Kalgash of almost the same mass as the planet. This moon would cause an eclipse of Dovim every 2049 years resulting in a period of absolute darkness, mass madness and the self-destruction of civilization.

## 3. OBSERVABLES

The following have been stated in the book (All page number refer to the 1991 edition (Asimov & Silverberg, 1991):

1. Kalgash orbits Onos which is a yellow star at a distance of 1.2 AU (pg. 185)
2. Trey and Patru are a close binary pair and appear white in color. They are faint and so must be white dwarfs (pg. 3)
3. Tano and Sitha are another close binary, also white in color at a distance of 13.2 AU from Onos. They are, brighter than Trey and Patru but still too faint to be main sequence stars and are also likely to be white dwarfs (pg. 6)
4. Dovim is a red star (pg. 78).
5. The furthest sun from Kalgash is 110 light-minutes away (13.4 AU), or somewhere just outside the orbit of Saturn as seen from the Sun (pg. 185)
6. The six suns of this system are the only stars in the Universe observable from Kalgash under normal conditions. Once every 2049 years, there is an eclipse of Dovim when no other of the suns are in the sky and the sky becomes dark enough to see the stars in the nearby cluster, contributing to the madness (pg. 111).
7. The Kalgash system is surrounded by a cluster of bright stars which are ordinarily, as mentioned above, not visible (pg. 215).
8. Astronomers at the beginning of the story discovered that Kalgash had a moon of about the same size. It eclipses Dovim every 2049 years at which time its angular diameter is seven times than that of the star. This results in a total eclipse and therefore darkness for a period of 9 - 14 hours, depending on latitude; long enough to ensure that every part of the planet experiences a period of darkness (pg. 114)

## 4. POSSIBLE SCENARIOS AND CONSEQUENCES

The most important constraint is that of the red star Dovim. This is the only star visible in that crucial time just before the deadly eclipse and the plot hinges on two factors:

1. Dovim is bright enough to blot out the other stars in the Universe.

2. The eclipse lasts for a period of 9 hours at the equator, sufficient that the entire planet is covered in darkness.

If we look at the case of our own Earth, the Sun has an apparent magnitude of -26.74 (Williams, 2004) and the faintest object visible to the naked eye during the day is -4.00 magnitudes (Wikipedia, 2014, http://en.wikipedia.org/wiki/Apparent_magnitude) for a difference of about 22.5 magnitudes. We will assume that Dovim is bright enough to blot out 0th magnitude stars and hence must be brighter than magnitude -22.5. This puts the star at a distance of about 44 light minutes from the Onos (Kalgash's primary star). The distance of the star from Kalgash would then vary from 34 to 54 light minutes and the star's angular diameter as observed from the planet would vary from 3 to 5 arcminutes.

The eclipse of Dovim by Kalgash 2 as seen from the planet lasts a minimum of 9 hours, compared to a solar eclipse on the Earth of less than 8 minutes (Meeus, 2003). Our Moon is about the same size as our Sun but Asimov does mention that the angular diameter of Kalgash 2 is seven times that of Dovim or about 2°. The orbital period of the Moon can be calculated from Kepler's Laws.

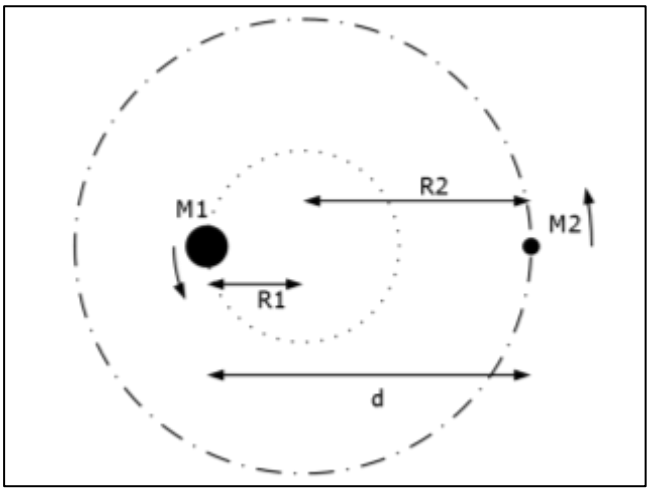

**Figure 1** Binary System in a circular Keplerian Orbit

We can derive the equation of motion of Kalgash and Kalgash 2 by equating the gravitational force to the centripetal force (we consider $M_1$):

$$\frac{M_1 v_1^2}{R_1} = \frac{G M_1 M_2}{(R_1 + R_2)^2} \quad (1)$$

$$\Rightarrow \frac{v_1^2}{R_1} = \frac{G M_2}{(R_1 + R_2)^2}$$

$$\Rightarrow \frac{\left(\frac{2\pi R_1}{T}\right)^2}{R_1} = \frac{G M_2}{(R_1 + R_2)^2}$$

$$\Rightarrow T^2 = \frac{4\pi^2 R_1 (R_1 + R_2)^2}{G M_2}$$

Since $M_1$ and $M_2$ orbit around their centre of mass, we can write:

$$R_1 = \frac{(R_1 + R_2) M_2}{M_1 + M_2} \quad (2)$$

We insert this result into (1):

$$T^2 = \frac{4\pi^2 (R_1 + R_2)^3}{G(M_1 + M_2)} = \frac{4\pi^2 d^3}{G(M_1 + M_2)} \quad (3)$$

$$\Rightarrow \boxed{T^2 = \frac{4\pi^2}{G(M_1 + M_2)} d^3}$$

We have considered a circular orbit, but the result is also valid for an ellipse. In that case, d is replaced with a, the semi-major axis. If $M_1 = M_2 = M$, then we get:

$$\boxed{T^2 = \frac{2\pi^2}{GM} a^3} \quad (4)$$

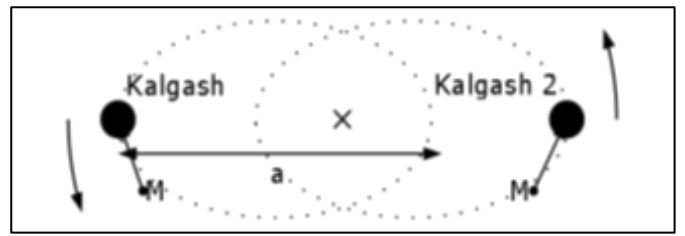

**Figure 2** Kalgash and its moon in an elliptical Keplerian Orbit

Now, we can calculate the angular speed of Kalgash 2, with respect to Kalgash:

$$\omega = \frac{1}{2}\sqrt{\frac{2GM}{a^3}} \tag{5}$$

During the eclipse, Kalgash 2 will cover a total angular distance of:

$$\theta = \theta_K + \theta_{K2} = 8.72 \cdot 10^{-3} \text{ rad} = \mathbf{26"} \tag{6}$$

Where '$\theta_K$' and '$\theta_{K2}$' are the angular diameters of Kalgash and Kalgash 2 respectively. Combining (5) and (6). We get the duration of the eclipse:

$$t = \frac{\theta}{\frac{1}{2}\sqrt{\frac{2GM}{a^3}}} \propto \left(\frac{a^3}{M}\right)^{0.5} \tag{7}$$

Next, we would like to eliminate the variables 'a' and 'M'. We note that the radius $R_{K2}$ of Kalgash 2 is:

$$R_{K2} = \frac{1}{2} \cdot (2a) \cdot \theta_{K2} \tag{8}$$

Then,

$$Density = \rho_{K2} = \frac{3M_{K2}}{4\pi(R_{K2})^3} \tag{9}$$

Inserting (9) into (7), we get:

$$t \propto \rho^{-0.5} \tag{10}$$

We summarise the results in the following graph (Figure 3) for a Kalgash 2 of Earth-mass.

We have a constraint on the size of Kalgash 2 from the stated fact that its angular diameter at the time of eclipse is 7 times that of Dovim. We have already assumed that the mass of the planet and the moon are the same as that of the Earth which thereby sets the diameter of the moon as a function of its density. If this density were the same as that of our own Moon, Kalgash 2 would be relatively small and close to the planet and therefore would move rapidly along its orbit resulting in short eclipses. On the other hand, if the density of Kalgash 2 were closer to that of the gas giants, the radius of the moon would be much larger and thus its distance from Kalgash would be much greater while maintaining the same angular diameter. Then by Kepler's Third Law, it would move more slowly along its orbit and the eclipse duration would be correspondingly longer. We find that a moon with a density the same as Saturn (~0.7 gm/cm³) (Williams, 2006) would eclipse Dovim for 9 hours, more than enough time to cover the entire planet with darkness. By comparison, the majority of the satellites in our Solar System have densities of 1.5 - 3 gm/cm³ (Williams, 2014) with the least dense large moon being Tethys with a density of 0.9 gm/cm³ (Roatsch, et al., 2009).

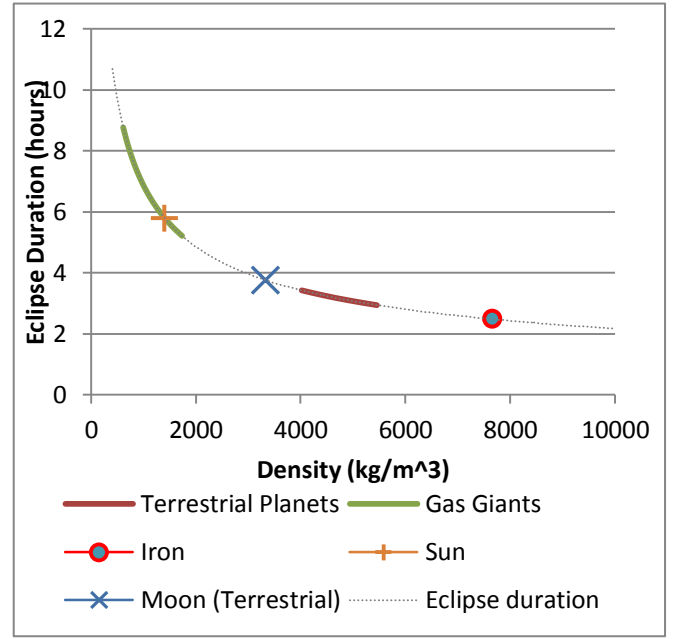

**Figure 3** Eclipse durations in function of density

# THE OTHER SUNS

## ONOS
Onos is the parent sun of Kalgash. It gives off a bright yellow light, therefore it must be a Sun-like star. However, it is slightly further away than our own sun, at a distance of 1.2AU. Onos would appear dimmer by a factor of 1.44, or like a bright day in Sao Paulo.

## TANO-SITHA
Tano-Sitha are also a white binary pair, further yet brighter than Trey - Patru. This indicates that they could be massive A-type stars or white dwarves.

*A-type Stars*
Since we want to minimise tidal effects, we shall suppose that they are A9V types at a temperature of 7100K, radius of 1.55 solar radii and 1.62 solar masses. This binary is at the edge of the Kalgash universe at a distance of 110 light minutes. It will have an apparent magnitude of -23.7, much brighter than Dovim! But Asimov mentions that they produce light much fainter than Dovim. They cannot be A-type Stars.

*White Dwarfs*
In this case, we suppose the stars are similar to Sirius B. They have a surface temperature of 25200K, a radius similar to that of Earth (6370km) and are at a distance of 110 light minutes. Then the apparent magnitude is -18.09 or about 114 times brighter than the full moon.

## TREY-PATRU

The secondary binary also produces white light. However, it is closer to Onos than Tano-Sitha, but fainter. The only way this is possible is if it is a cool white dwarf, similar to 40 Eridani B. Then the stars will have a temperature of 16500K, a radius of 0.014 Solar radius and a mass of 0.50 solar mass. At a distance of 100 light-minutes, the apparent magnitude of the binary is -17.38, or around 59 times brighter than the moon.

# SIMULATION

We have written a program computer program that simulates a stellar system. *We have not attempted to match the scale of the system, this is only a demonstration*. The user inputs the initial positions and velocities of the stars, as well as a precision parameter which determines how often forces and velocities of the individual bodies are updated and recalculated. The simulation is based on Newton's laws of motion:

1. $r = v * dt + r_0$
2. $v = a * dt + v_0$
3. $a = \sum_i \frac{GM_i}{(\Delta R_i)^2} \overrightarrow{\Delta R_i}$

where 'r' and '$r_0$' are the instantaneous and initial positions, 'v' and '$v_0$' are the instantaneous and initial velocities, 'a' is the instantaneous acceleration and 'dt'[1] is the time step in seconds. The smaller the value of dt, the better the simulation but the longer the simulation length.

## KALGASH 2 DISTANCE

We will suppose that Kalgash 2 is a gaseous moon slightly less dense than Saturn (0.687g/cm$^3$). Since it has the same mass as the earth, it will have a diameter of 26,540 km. It also subtends an angle of 26". This means that Kalgash 2 is at a distance of 3.04 million kilometres. However, both the planet and the moon are light compared to the stars and we can ignore them in the simulation.

|  | Mass (kg) | Distance from Kalgash (m) |
|---|---|---|
| Onos | 2.00*10^30 | 1.80*10^11 |
| Dovim | 1.24*10^30 | 7.91*10^11 |
| Tano-Sitha | 3.92*10^30 | 1.98*10^12 |
| Trey-Patru | 2.00*10^30 | 1.80*10^12 |
| Kalgash | 6.00*10^24 | 0 |
| Kalgash 2 | 6.00*10^24 | 3.04*10^6 |

**Figure 4** Distances to the Suns, Planets and Moons

As discussed above, the stellar system of Asimov's Nightfall has 6 stars but with two sets of binaries (Trey-Patru and Tano-Sitha), which we treat as single point masses. Although different stable configurations may be possible, we have worked out one possible configuration[2] which we found to be stable for a few hundred years. We further arrange these stars into three sets of 2-star systems. First, Onos and Dovim (circling each other) as one independent system, then Tano-Sitha and Onos-Dovim (Figure 5), where the forces act on the centre of masses of the two binaries. And finally, Trey-Patru and Onos-Dovim-Tano-Sitha.

---

[1] The value of dt is critical in the simulations. We have used 1000sec. This means that the errors will become significant over large periods of time. The simulations here just give an approximate idea of the system stability.

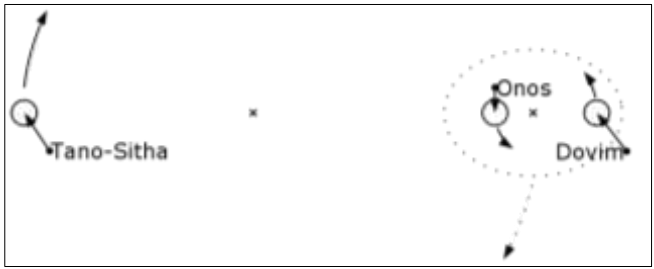

**Figure 5** Tano-Sitha orbiting Onos-Dovim

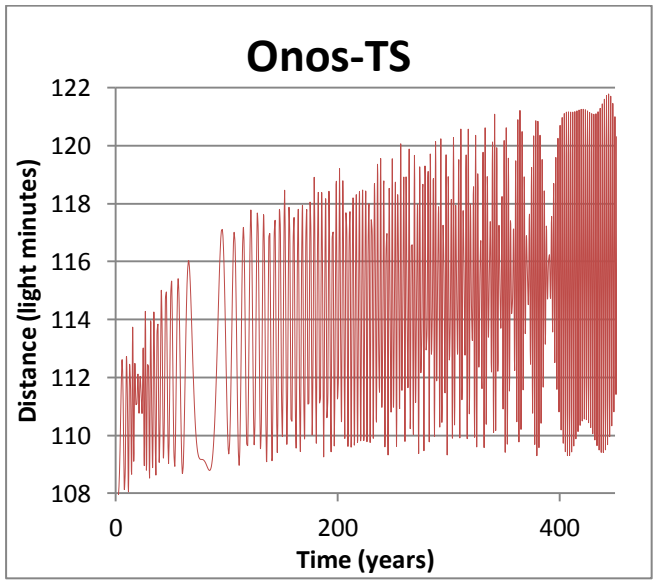

**Figure 6** Distance between Onos and Tano-Sitha in function of time for the configuration in figure 5.

The graph is similar to a velocity graph of a binary star (in this case Onos-Dovim is the binary, since we have treated TS as a single star). The amplitudes of the oscillation slowly increase, but this is within the limits of the precision settings of our program. Also the amplitudes increase over a period of several hundred years: the people of Kalgash need have no fear of being swallowed by a sun!

---

[2] In this configuration we have changed some distances to allow a stable configuration.

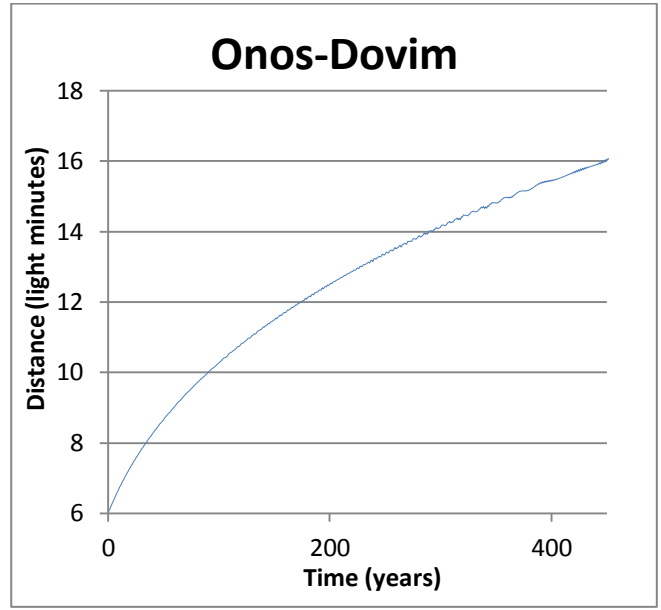

**Figure 7** Distance between Onos and Dovim in function of time for the configuration in figure 5.

Next, we add Trey-Patru. Now we suspect that the addition of Trey-Patru will quickly rip the system. To counter this, we make sure that Trey-Patru is not close to Onos-Dovim for large periods of time. This is done by giving TP an opposite angular velocity as the following image shows:

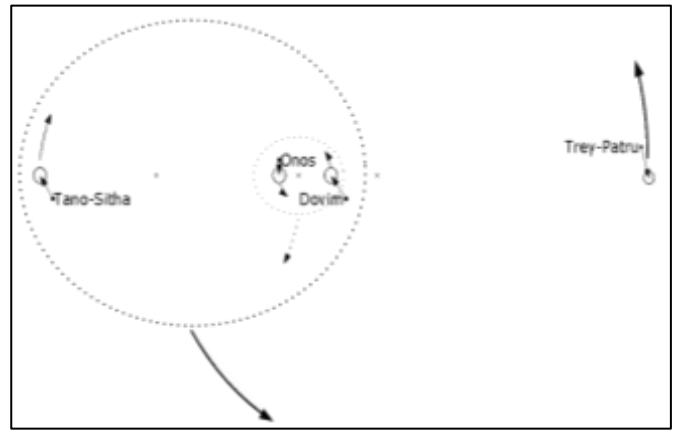

**Figure 8** Trey-Patru orbiting Tano-Sitha-Onos-Dovim.

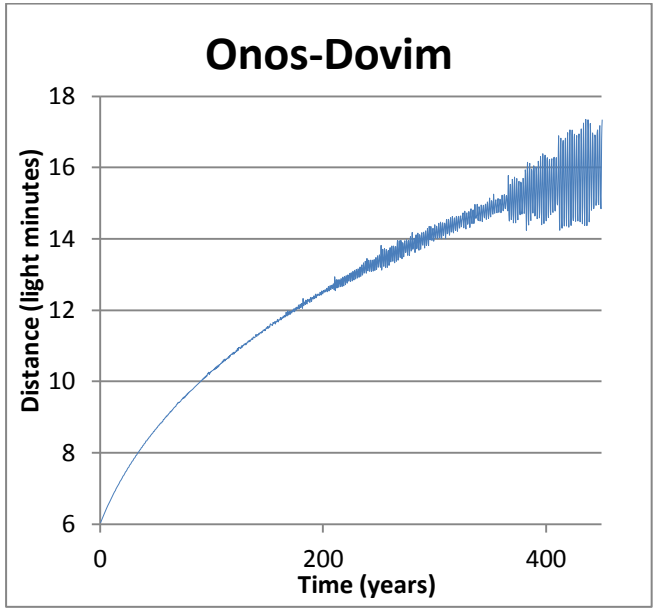

**Figure 9** Distance between Onos and Dovim in function of time for the configuration in figure 8.

The addition of TP increases the tidal forces on Onos-Dovim. However, for the duration of the story (about 2 years) these effects will not be perceptible. (Though subsequent generations of the Kalgash people will face dangerous scenarios.)

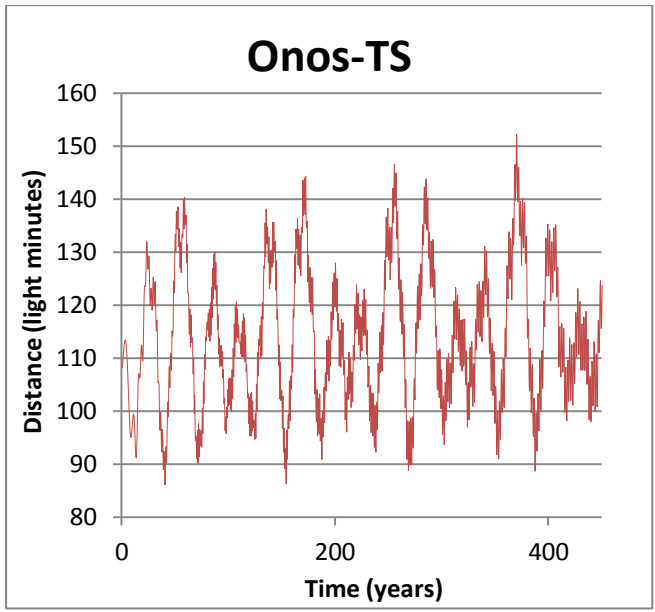

**Figure 10** Distance between Onos and Tano-Sitha in function of time for the configuration in figure 8.

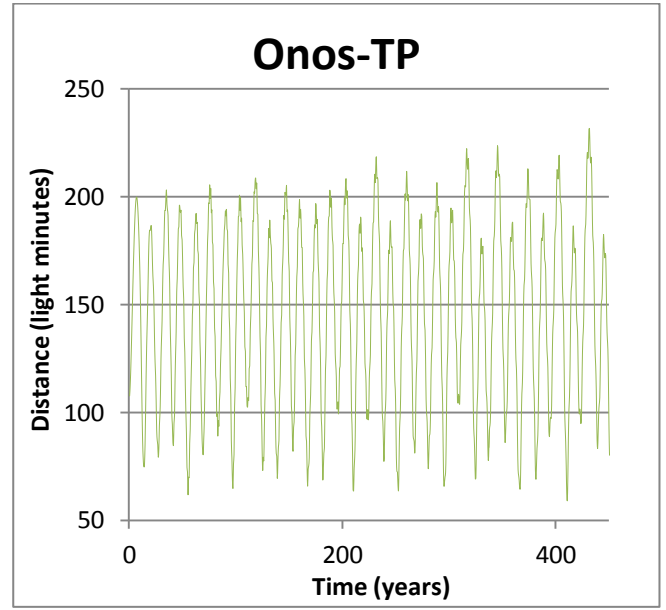

**Figure 11** Distance between Onos and Trey-Patru in function of time for the configuration in figure 8.

The Kalgash people measured the universe to be 110 light minutes across. Perhaps, they made this measurement at a very specific time in the evolution of the system.

Luckily for them, the hottest binary star maintains a safe distance from them. The large amplitudes mean that at some later time, the "stars" might be become visible.

## CONCLUSION

We have explored several aspects of Asimov's novel. We have found that the suns, especially Dovim are bright enough to blot out the stars. Kalgash 2 can eclipse Dovim for a period of 9 hours.

We also tested one possible star configuration and after running some simulations, we found that the system is possible for short periods of time. Several other configurations might exist and have to be explored more fully.

# REFERENCES

Asimov, I., 1941. *Nightfall.* [Online]
Available at:
https://www.uni.edu/morgans/astro/course/nightfall.pdf
[Accessed April 2014].

Asimov, I. & Silverberg, R., 1991. *Nightfall.* New York, NY: Bantam Books.

Jens Buus, J. M., n.d. *Latitude Dependence of the Maximum Duration of a total Solar Eclipse.* [Online]
Available at: www.eclipse-chasers.com/papers/Maximum_duration.pdf
[Accessed 11 July 2014].

Meeus, J., 2003. The Maximum Possible Duration of a Total Solar Eclipse. *Journal of the British Astronomical Association,* 113(6), pp. 343-348.

Orosz, J. A. et al., 2012. *Sciencemag.* [Online]
Available at:
https://www.sciencemag.org/content/337/6101/1511.short
[Accessed May 2014].

Roatsch, T., Jaumann, R. & Thomas, P., 2009. *Wikipedia - Cartographic Mapping of the Icy Satellites Using ISS and VIMS Data.* [Online]
Available at:
http://en.wikipedia.org/wiki/Tethys_%28moon%29
[Accessed 11 July 2014].

Sagan, C., 1992. *Nature.* [Online]
Available at:
http://www.nature.com/nature/journal/v357/n6374/pdf/357113a0.pdf
[Accessed May 2014].

Wikipedia, 2014. *Apparent Magnitude.* [Online]
Available at:
http://en.wikipedia.org/wiki/Apparent_magnitude
[Accessed July 2014].

Williams, D. D. R., 2004. *Wikipedia.* [Online]
Available at:
http://en.wikipedia.org/wiki/Apparent_magnitude
[Accessed May 2014].

Williams, D. D. R., 2014. *Nasa.* [Online]
Available at:
http://nssdc.gsfc.nasa.gov/planetary/factsheet/index.html
[Accessed 11 July 2014].

Williams, D. R., 2006. *Wikipedia.* [Online]
Available at: http://en.wikipedia.org/wiki/Saturn
[Accessed 11 July 2014].